\documentclass[aps,pra,9pt,twocolumn,showpacs,superscriptaddress]{revtex4-2}
\usepackage{physics}							
\usepackage{latexsym}
\usepackage{amssymb}
\usepackage{graphics,epstopdf}
\usepackage{newlfont}
\usepackage{amsfonts}
\usepackage{epsfig}
\usepackage[colorlinks=true, citecolor=blue, urlcolor=blue]{hyperref}
\usepackage{amsmath}
\usepackage{graphicx}
\usepackage{dcolumn}
\usepackage{bm}
\usepackage{color}
\usepackage{longtable}
\usepackage{amsthm}
\usepackage{times}
\usepackage{mathtools}

\newtheorem*{theorem*}{Theorem}

\begin{document}
 \title{\large  Remote Creation of Quantum Coherence via Indefinite Causal Order}

\author{Jasleen Kaur}
\email{mailjkaur7@gmail.com}
\affiliation{Department of Applied Physics, Amity Institute of Applied Sciences, Amity University, Noida 201 313, India}
\author{Shrobona Bagchi (corresponding author)}
\email{shrobonab@mail.tau.ac.il}
\affiliation{Raymond and Beverly Sackler School of Physics and Astronomy, Tel Aviv University, Tel-Aviv 69978, Israel}
\author{Arun K Pati}
\email{akpati@hri.res.in}
\affiliation{Quantum Information and Computation Group, Harish-Chandra Research Institute, HBNI, Allahabad, India}
\affiliation{Homi Bhabha National Institute, Training School Complex, Anushakti Nagar, Mumbai 400085, India.}

\maketitle

\section {Abstract }

 Quantum coherence is a prime resource in quantum computing and quantum communication. Quantum coherence of an arbitrary qubit can be created at a remote 
 location using maximally entangled state, local operation and classical communication. However, if there is a noisy channel acting on one side of the shared resource, then, it is not possible to create perfect quantum coherence remotely. Here, we present a method for the creation of quantum coherence at a remote location via the use of entangled state and indefinite causal order. We show this specifically for the 
 superposition of two completely depolarizing channels, two partially depolarizing channels and one completely depolarizing channel along with a unitary operator. We find that when the indefinite causal order of channels act on one-half of the entangled pair, then the shared state looses entanglement, but can retain non-zero quantum discord. This finding may have some interesting applications on its own where discord can be consumed as a resource. Our results suggest that the indefinite causal order along with a tiny amount of quantum discord can 
 act as a resource in creating non-zero quantum coherence in the absence of entanglement.

\section {Introduction }

One of the main feature that differentiates the quantum world from the classical world is the superposition principle for 
quantum states. It gives rise to many counter-intuitive
features such as the uncertainty relations \cite{H,rob,sch,mac}, quantum entanglement \cite{Horo} and quantum coherence \cite{Alex}. 
Recent developments show that quantum coherence plays an important role in the basic laws of quantum thermodynamics \cite{t1,t2,t3,t4,t5,t6,t7,t8,t9,t10}, nanoparticle systems, complex biological systems \cite{n1,n2,b1,b2,b3,b4,b5} and in making quantum states magical \cite{chira}. Even the resource theoretic quantification of quantum coherence has been established in Ref.\cite{Alex}.  It has found its usefulness in different areas of quantum technologies, quantum algorithms \cite{mark,fan,namit}, quantum metrology \cite{metro} and quantum channel discrimination \cite{chd}. 

Yet another very important and useful resource in quantum information processing tasks is the quantum entanglement. It arises as a result of superposition principle and the tensor product structure of the Hilbert spaces when we have two or more particles. Quantum entanglement
has been shown to be widely useful in numerous quantum technologies, quantum metrology, quantum thermodynamics \cite{Horo}. The most prominent use of quantum entanglement is in the 
context of quantum teleportation \cite{bennett} and the remote state preparation \cite{pati}. In quantum teleportation, one transfers an unknown quantum state to a distant party with the help of entangled state, local operations and classical communication. On the other hand, in remote state preparation \cite{pati}, one can prepare a known quantum state by using an entangled state along with local operation and classical communication.

Recently, the principle of superposition has been extended to the case of physical processes and this has been very useful, specifically, in aiding the quantum communication and metrological tasks.
The indefinite causal order \cite{i1,i2,i3,i4,i5} or the quantum SWITCH formalism, which is nothing but the superposition of causal order of quantum operations aided by a control qubit is one type of such superposition. There are other examples such as those generated by coherent control. The quantum SWITCH can give rise to quantum advantages in quantum computing \cite{comp}, reducing the communication complexity \cite{compl1,compl2}, quantum communication\cite{comm}, quantum teleportation \cite{chira1}, as metrological resource for quantum thermometry \cite{chira2} and in the task of channel discrimination \cite{chd1,chd2}. It is not only a theoretical possibility, but has also been experimentally realized \cite{exp1,exp2}.
Therefore, motivated by the above tasks that use entanglement and quantum superposition of channels as resources, we ask the question whether we can create quantum coherence at a distant location by using an entangled state, local operation and classical communication aided by the indefinite causal order. In this article, we find conditions for which this is possible. To actuate the scene, imagine that Alice can implement coherent operations in her lab and hence, can create quantum coherence starting from some incoherent state. But, Bob on the other hand, can implement only incoherent operations in his lab and thus cannot create quantum coherence which he may like to use later for some practical purposes. However, if Alice and Bob share an entangled state, then, Alice can help to prepare quantum coherence at a remote location using entangled state, local operation and classical communication. But, it is not possible for Alice to perfectly prepare quantum coherence remotely, if there is noisy channel acting on one side of the shared resource. Here, we discuss several methods for the creation of non-zero quantum coherence at a remote location via the use of entangled state and indefinite causal order. We find that when the indefinite causal order of channels act on one-half of the entangled pair, then, the shared state looses entanglement completely. Nevertheless, we find that the shared state can retain non-zero quantum discord. This finding may have some interesting applications where discord can be consumed as a resource in quantum communication. Our results suggest that the indefinite causal order can act as a resource in creating non-zero quantum coherence in the absence of entanglement, in the presence of a small amount of quantum discord.
 
The paper is organized as follows. In section II, we state necessary background needed for our analysis which includes the theory of quantum coherence and indefinite causal order. In section III, we discuss the basic formalism for remote creation of quantum coherence. In section IV, we present explicit protocols to state our main results which include the remote creation of quantum coherence via indefinite causal order when we have two completely depolarizing channel, one depolarizing and one unitary and two partially depolarizing channels acting on one half of the maximally entangled state. In section V, we provide a summary of our results.

\section {Background }

\subsection{Quantum coherence}
The notion of quantum coherence by definition is a basis dependent quantity. Therefore, we fix the basis at first which is called the reference basis. Let us denote an orthonormal basis $|i\rangle$  $( i = 0, 1, \ldots d-1)$ in a $d$ dimensional Hilbert space ${\cal H}^d$. 
 All the states that are diagonal in this basis are the incoherent states, i.e., the states of the form $ \sum_i p_i|i\rangle\langle i|$ with $\sum_i p_i =1$ have zero coherence in the chosen basis. Thus, all the states which have non-zero off-diagonal elements in the reference basis form the 
coherent states and these states will have a finite value of coherence. The quantification in terms of the off-diagonal elements in a quantum state in the reference basis gives a measure of quantum coherence. For our purpose
we will use the $l_1$ norm to quantify quantum coherence. For a $d$ dimensional pure state 
$|\psi\rangle=\sum_i c_i|i\rangle \in {\cal H}^d $, 
the $l_1$ norm coherence is defined as $ C_{l_1}(\psi) = \sum_{i\neq j}|c_i| | c_j|$. Quantum coherence satisfies some reasonable properties that we discuss here \cite{a,b,lm}. First, it is non-negative, i.e., $C(\rho)\geq 0$ with equality if and only if $\rho$ is an incoherent state. Secondly, monotonic under any incoherent operations, which is $C(\Lambda(\rho))\leq C(\rho)$ for any incoherent operation $\Lambda$. It is strongly monotonic, i.e., it does not increase on an average under selective incoherent operations, which is $\sum_iq_iC(\sigma_i)\leq C(\rho)$, with probabilities $q_i=\mathrm{Tr}[K_i\rho K_i^\dagger]$,
post-measurement states $\sigma_i=\frac{K_i\rho K_i^\dagger}{q_i}$ and incoherent operators $K_i$. Also it is a convex function of the state, i.e., for a density operator 
$\rho= \sum_i p_i \rho_i$, we have $\sum_i p_i C(\rho_i)\geq C(\sum_i p_i\rho_i)$. The $l_1$ norm measure of quantum coherence satisfies all these properties. Specifically, if we have a mixed state 
$\rho=\sum_i p_i \rho_i$, then the $l_1$ norm quantum coherence of $\rho$ is given by $\sum_{i\neq j}|\rho_{ij}|$, , where $\rho_{ij}$ is the $\{i,j\}^{th}$ element of the density matrix $\rho$. The $l_1$ norm of coherence has an operational meaning as it captures the wave nature of the quantum system and satisfies a duality relation between the path information \cite{bera}, thus capturing the wave aspect of the quantum entity.

\subsection{Indefinite causal order}

In the classical world,  we are predisposed to think of events occurring in a fixed causal order. However it has been shown in quantum mechanics that one can have non-classical causal structures in which the order of events is not definite \cite{chiri,brukner}. It  has  been  shown that  such an indefinite causal order provides an advantage in
certain quantum information processing tasks \cite{chiri2,col,ebler}, quantum computation \cite{ara}, and even in quantum communication complexity \cite{Feix, Guerin}. In practice, one can implement an operation with indefinite causal order via the SWITCH formalism. In this formalism, there is superposition of causal orders of two or more quantum operations or
quantum channels aided by a control qubit. If we have two operations ${\cal M}_1$ and ${\cal M}_2$, then a state can be operated either by ${\cal M}_1  {\cal M}_2$ or ${\cal M}_2  {\cal M}_1$ in the causal setting.
That is to say, it can go through ${\cal M}_1 {\cal M}_2$ when the control state is $|0\rangle\langle 0|$ or it 
can go through ${\cal M}_2 {\cal M}_1$ when the control qubit is $|1\rangle\langle 1|$.
But with the help of a control qubit $\frac{1}{\sqrt{2}}(|0\rangle +|1\rangle)$,  the input state can go through both the configurations at the same time, leading to superposition of causal orders. 
Now let us denote the Kraus operators of ${\cal M}_1$ by $\{X_i\}$ and those of ${\cal M}_2$ by $\{Y_i\}$. Then,
the Kraus operator for the superposition of quantum channels is given by $W_{ij}=|0\rangle\langle 0|\otimes X_iY_j +|1\rangle\langle 1|\otimes Y_jX_i$. Therefore, when the SWICTH acts on the input state, we get the 
output state as the following 
\begin{align}\nonumber
&\rho\otimes\rho_c\rightarrow \\ 
&\rho_{\mbox{out}}=S({\cal M}_1, {\cal M}_2)(\rho\otimes\rho_c)=\sum_{ij}W_{ij}(\rho\otimes\rho_c)W_{ij}^\dagger,
\end{align}
where $\sum W_{ij}^\dagger W_{ij}=\mathbb{I}$. The above formula represents the action of quantum SWITCH that implements the superposition of causal order of two quantum channels via a control qubit.

\section{Remote creation of quantum coherence}

Here, we introduce the basic idea behind the remote creation of quantum coherence \cite{akp21}. Let Alice and Bob share an entangled state $\rho_{12}$. Imagine that Alice has a state her mind, i.e., she has complete 
knowledge about a state $\rho = \ket{\psi} \bra{\psi}$. Alice will perform a local unitary operation depending on the knowledge she has about  the qubit. Then, the shared state $\rho_{12}$ 
will undergo a transformation $\rho_{12} \rightarrow (U \otimes I) \rho_{12} (U^{\dagger} \otimes I) = {\tilde \rho}_{12}$. She will perform a projective measurement in the computational basis $\{\ket{0}, \ket{1} \}$.
As a result of this local measurement, the state will undergo a transformation as given by
\begin{align}
{\tilde \rho}_{12} \rightarrow {\tilde \rho}_{12}^{(0)} = \frac{ (\Pi_0 \otimes I ){\tilde \rho}_{12} (\Pi_0 \otimes I )}{\Tr [(\Pi_0 \otimes I ){\tilde \rho}_{12} (\Pi_0 \otimes I )] },\nonumber \\
{\tilde \rho}_{12} \rightarrow {\tilde \rho}_{12}^{(1)} = \frac{ (\Pi_1 \otimes I ){\tilde \rho}_{12} (\Pi_1 \otimes I )}{\Tr [(\Pi_1 \otimes I ){\tilde \rho}_{12} (\Pi_1 \otimes I )] },
\end{align}
where $\Pi_0=\ket{0}\bra{0}$ and $\Pi_1=\ket{1}\bra{1}$. Depending on the measurement outcome, she will communicate the classical information to Bob. After receiving the classical communication, Bob may or may not apply a local unitary, depending on the protocol. 
At the end, the state of the qubit at Bob's location will have the desired amount of coherence that Alice wanted to create. 
That is to say that at Bob's location, we have $C(\rho) = C({\tilde \rho}_{2}^{(0)} ) = C({\tilde \rho}_{2}^{(1)} )$, where ${\tilde \rho}_{2}^{(0)}= \Tr \tilde {\rho}_{12}^{(0)}$ and 
$\tilde {\rho}_{2}^{(1)}= \Tr \tilde {\rho}_{12}^{(1)}$.

To provide a concrete protocol, let us consider the scenario where Alice has a known state of a qubit $\ket{\psi}=\alpha \ket{0}+ \beta \ket{1} \in {\cal H}^2$ in her lab. The amount of coherence for the state $\ket{\psi}$ is $ C(\psi) = 2 |\alpha||\beta|$. 
She wants to create a state in Bob's lab whose coherence will be the 
same as the that of the known state in her lab. Let us imagine that Alice and Bob shares an EPR pair $\ket{\Psi^{-}}=\frac{1}{\sqrt{2}}(\ket{0} \ket{1} -\ket{1} \ket{0} )$. The state $\ket{\Psi^{-}}$ can be written in the following form 
 \begin{align}
 \ket{\Psi^{-}}=\frac{1}{\sqrt{2}}(\ket{\psi}\ket{\Bar{\psi}}-\ket{\Bar{\psi}}\ket{\psi}),
 \end{align}
  where $\ket{\psi}=\alpha \ket{0}+ \beta \ket{1}$ is an arbitrary state and $|\bar{\psi}\rangle = \alpha^*|1\rangle -\beta^*|0\rangle$ is its orthogonal state. For a given state $\ket{\psi}=\alpha \ket{0}+ \beta \ket{1}$ let us define a unitary operator in  the following way
   \begin{align*}
U(\alpha,\beta) = \begin{bmatrix}
	\alpha & -\beta^* \\
  	\beta & \alpha^*
	\end{bmatrix},
\end{align*}
 where $U(\alpha,\beta)\ket{0}=\ket{\psi}$ and $U(\alpha,\beta)\ket{1}=\ket{\bar{\psi}}$.
Let Alice applies $U^\dagger(\alpha,\beta)$ on her part of the shared entangled state $\ket{\psi^-}$ and Bob does nothing on his part. After Alice applies this operation, the shared entangled state can be expressed as
\begin{align}
U^\dagger(\alpha,\beta)\otimes \mathbb{I} \ket{\Psi^-}
=\frac{1}{\sqrt{2}}(\ket{0}\ket{\Bar{\psi}}-\ket{1}\ket{\psi}).
\end{align}
  Now, Alice does a projective measurement in the basis $\{\ket{0}, \ket{1} \}$. Let these projectors be denoted as $\Pi_0=\ket{0}\bra{0}$ and $\Pi_1=\ket{1}\bra{1}$. If $\Pi_0$ clicks, after Alice communicates to Bob, the state at Bob's lab is $\ket{\bar{\psi}}$.  If $\Pi_1$ clicks, after Alice communicates the measurement outcome to Bob, 
  the state at Bob's lab is $\ket{\psi}$. If $\Pi_1$ clicks, then Bob does nothing and the desired coherence is created at remote location.  If $\Pi_0$ clicks, then Bob can apply $i \sigma_y$ and this 
  local operation can convert $\ket {\bar \psi}$ to $\ket{\psi^*}$, where $\ket{\psi^*}= \alpha \ket{0}+ \beta^* \ket{1}$ is the conjugated ket. Since $\ket{\psi}$ and $\ket{\psi^*}$ have 
  same amount of coherence,  i.e., 
  $ C(\psi) = C(\psi^* ) = 2|\alpha||\beta|$, we can say that 
  the desired coherence is created at remote location. Therefore, with the help of the shared maximally entangled state and one bit of classical information about her measurement outcome, 
  Alice is able to prepare the coherence of any known pure quantum state $\ket{\psi}$ at a distant location. 
  One can also simplify the protocol by allowing Alice to perform measurement directly using $\{ \ket{\psi}, \ket{ \bar{\psi} } \}$ basis and then communicating the measurement outcome to Bob. As Alice has complete knowledge about the state of the qubit, she can perform the aforesaid measurement, in principle. In the second version of our protocol, we notice again that 
 $|\psi\rangle$ and $|\bar{\psi}\rangle$ have the same $l_1$ norm of coherence, i.e.,  $ C(\psi) = C(\bar{\psi} ) = 2|\alpha||\beta|$.
  Though, this scheme looks simple, this cannot be generalised to higher dimensions \cite{akp21}. It is still an open question. 
  
  Now, if there is a noisy channel acting on one side of the entangled pair, then the perfect remote creation of quantum coherence of a qubit is not possible. This is the main focus of the present paper. In the sequel, we discuss several scenarios, where the use of indefinite causal order can help in creating non-zero quantum coherence 
  even in the presence of noise.

\section{Creation of Coherence with Indefinite Causal Order}

Let Alice and Bob share a maximally entangled state. In general, if half of the shared state undergoes a noisy channel, then the shared state may or may not retain entanglement and coherence.  For example, if we apply completely depolarising channel, then the shared state is not entangled anymore and completely decoheres too. But under the action of some other channels, it may loose some amount of entanglement or coherence. Since single use of channel is not able to retain full coherence, we investigate several scenarios where channels are under superposition of indefinite causal order. In this respect, we know that a random mixture of non-commuting processes can effectively work as a  noisy channel. Here, we take the following noisy scenarios into consideration.

\subsection{Two completely depolarizing channels}

This case deals with the self-switching phenomenon. The initial state shared by Alice and Bob is taken to be the EPR pair which is $|\Psi^-\rangle=\frac{1}{\sqrt{2}}(|01\rangle-|10\rangle)$. The noise comprising of the two causally activated depolarizing channels acts on half of the EPR pair possessed by Bob. Therefore, the state shared between Alice and Bob after the action of the overall quantum channel resulting from the quantum SWITCH of two completely depolarizing channels is given by
\begin{align}
  \rho^{out}_{12c}=\sum_{ij}(\mathbb{I}\otimes W_{ij})(|\Psi^-\rangle\langle\Psi^-|\otimes \rho_c)(\mathbb{I}\otimes W_{ij}^{\dagger}),
 \end{align}
where $W_{ij}$ will act on half of the EPR pair and the control qubit. In the above equation we have
\begin{align}\nonumber
\rho_c=|\Psi_c\rangle\langle\Psi_c|~ \mathrm{where}~|\Psi_c\rangle=\sqrt{p}|0\rangle+\sqrt{1-p}|1\rangle,\\ ~~\mathrm{and} ~~W_{ij}=\frac{1}{4}(\sigma_i\sigma_j\otimes |0\rangle\langle 0|+ \sigma_j\sigma_i\otimes|1\rangle\langle 1| ).
 \end{align}
Simplifying the above equation and denoting $k=\sqrt{p(1-p)}$ with $0\leq k\leq 1$, we obtain the following expression for the joint state of 
two qubits and the control qubit as given by

\begin{align}\nonumber
  \rho^{out}_{12c}=\frac{\mathbb{I}}{2}\otimes \frac{\mathbb{I}}{2} \otimes \{ p\ket{0}\bra{0}+ (1-p)\ket{1}\bra{1}\}
   + \frac{1}{2}\{ \ket{0}\bra{0}\\ \nonumber \otimes\ket{1}\bra{1}+  \ket{1}\bra{1}\otimes\ket{0}\bra{0}\}  \otimes \frac{k}{4}\{ \ket{0}\bra{1}+\ket{1}\bra{0}\} \\ \nonumber
    - \frac{k}{16}\{ \ket{0}\bra{1} \otimes (\sigma_x-i\sigma_y)\otimes(\ket{0}\bra{1}+\ket{1}\bra{0}) \} \\ -\frac{k}{16}\{ \ket{1}\bra{0} \otimes (\sigma_x+i\sigma_y)\otimes(\ket{0}\bra{1}+\ket{1}\bra{0})\}.
\end{align}
Now, if measurement if performed on the control qubit in the $\{ \ket{+}, \ket{-}\}$ basis, then the conditional state shared between Alice and Bob (for the measurement outcome, for e.g., $\ket{+}$ ) is given by 
\begin{equation}
    \begin{split}
        \rho_{12}^+= \frac{1}{2}\{ \frac{\mathbb{I}}{2}\otimes \frac{\mathbb{I}}{2}\} + \frac{k}{8}\{ \ket{0}\bra{0}\otimes\ket{1}\bra{1}+ \ket{1}\bra{1}\otimes\ket{0}\bra{0}\} \\ - \frac{k}{16}\{\ket{0}\bra{1}\otimes(\sigma_x-i\sigma_y)+ \ket{1}\bra{0}\otimes (\sigma_x
        +i\sigma_y)\}\\ 
        = \frac{1}{2}\{ \frac{\mathbb{I}}{2}\otimes \frac{\mathbb{I}}{2}\} + \frac{k}{8}\{
\ket{0}\bra{0}\otimes\ket{1}\bra{1} + \ket{1}\bra{1}\otimes\ket{0}\bra{0}\\  - \ket{0}\bra{1}\otimes\ket{1}\bra{0}- \ket{1}\bra{0}\otimes\ket{0}\bra{1}\}\end{split}  
\end{equation}    
One may ask, whether the superposition of quantum channels can preserve some entanglement for the state $\rho_{12}^+$? To answer this question, 
we check if the shared state between Alice and Bob is still entangled by means of a partial transpose operation on the above state with respect to the first party. This is given by
\begin{equation}
  \rho_{12}^{T_A}=c \begin{bmatrix}\frac{1}{8} & 0 & 0 & \frac{-k}{8} 
    \\0 & \frac{1+k}{8} & 0 & 0 
    \\ 0 & 0 & \frac{1+k}{8} & 0
    \\ \frac{-k}{8} & 0 & 0 & \frac{1}{8},
    \end{bmatrix}  
\end{equation}
where $c= \frac{4}{2+ k} $ and $\rho_{12}^{T_A}$ is the partial transpositoned operator of $\rho_{12}^+$. Since $ \rho_{12}^{T_A}$ has three eigenvalues given by $\frac{1}{8}(1+k)$ and the fourth one given by $\frac{1}{8}(1-k)$, which are all positive, it can be stated using the PPT criterion of entanglement that the state which was initially maximally entangled has become separable under the noisy scenario. Since the conditional state has no entanglement, we consider if there can be some other quantum correlation shared between Alice and Bob ? For this purpose, we check the quantum discord in the state $\rho_{12}^+$ and find that the state has very small amount of quantum discord. Quantum discord is a kind of quantum correlation which is different than quantum entanglement. It is given by the difference in the total correlation measure called the quantum mutual information and the classical correlation measure given by the classical mutual information. Thus, the quantum discord is given by the following expression \cite{zurek} 
\begin{align}
    D(B|A)=I(A:B)-\mathrm{max}_{\Pi_{A_i}}J(A:B),
\end{align}
where the first term corresponds to quantum mutual information given by $I(A:B)=H(\rho_A)+H(\rho_B)-H(\rho_{AB})$, $H(.)$ being the von-Neumann entropy, and the second term on the right hand side corresponds to the classical correlation given by $H(B)-\mathrm{min}_{\Pi_{A_i}}H(B|A)$ where projective measurements are performed on the subsystem  $A$ \cite{vedral}. Typically, the calculation of quantum discord involves optimisation over 
the measurement basis and it is not easy to do analytically. However, in the present case, the calculation of quantum discord can be done analytically. Since the state is a $X$-state with maximally mixed marginals, and there exists closed form expression of quantum discord for states with maximally mixed marginals \cite{luo,xs}, the quantum discord is given by the following formula 
\begin{align}\nonumber
    D(\rho_{12}^+)=\frac{1}{4}[(1-c_1-c_2-c_3)\log_2(1-c_1-c_2-c_3)\\ \nonumber 
    +(1-c_1-c_2-c_3)\log_2(1-c_1-c_2-c_3)\\\nonumber +(1-c_1-c_2-c_3)\log_2(1-c_1-c_2-c_3)\\ \nonumber+(1-c_1-c_2-c_3)\log_2(1-c_1-c_2-c_3)]\\ \nonumber-\frac{1-c}{2}\log_2(1-c)-\frac{1+c}{2}\log_2(1+c),
\end{align}
where we have 
\begin{align}
    c_i=\mathrm{Tr}(\sigma_i\otimes\sigma_i\rho_{12}^+),~~~~~~c=\mathrm{max}\{|c_1|,|c_2|,|c_3|\},
\end{align}
with $\sigma_i$ being the Pauli matrices. Therefore, we have the following values for $c_i$s as given by  
\begin{align}
c_1=c_2=c_3=-\frac{k}{2+k}~\mathrm{and}~ c=\frac{k}{2+k}.
\end{align}
Using the above, expression for the quantum discord is given by
\begin{align}\nonumber
D(\rho_{12}^+)=\frac{1}{4}[3(1-\frac{k}{2+k})\log_2(1-\frac{k}{2+k})\\ \nonumber
+(1+\frac{3k}{2+k})\log_2(1+\frac{3k}{2+k})]\\ \nonumber
-(1-\frac{k}{2+k})\log_2(1-\frac{k}{2+k})\\
-(1+\frac{k}{2+k})\log_2(1+\frac{k}{2+k}).
\end{align}
 \begin{figure}[h]
  \centering
  \includegraphics[width=0.995
  \linewidth]{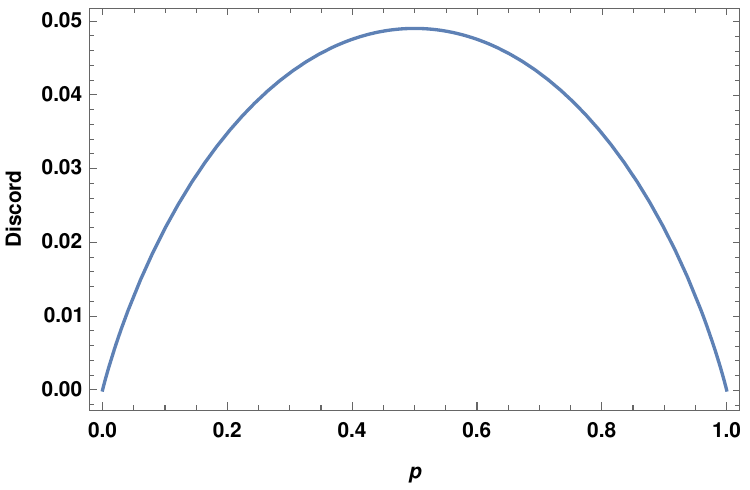}
  \caption{Quantum discord of the conditional state $\rho_{12}^+$ shared between Alice and Bob.}
  \label{fig::dis}
\end{figure}
We plot this as a function of $p$ in Fig.(\ref{fig::dis}) which shows a very small value of quantum discord. The maximum of value of quantum discord reaches $0.05$. Interestingly, we find that the quantum discord reaches its maximum values when the control qubit is in a maximally coherent state, i,e., an equal superposition of computational basis states. This indicates that a maximally coherent control qubit enables to maintain a tiny amount of quantum discord that is shared between Alice and Bob. To create quantum coherence at a remote location, let Alice  performs measurement in the basis $\{\ket{\psi}, \ket{ \bar {\psi}}\}$. After performing the conditional measurement on the control qubit, the shared state between Alice and Bob is given by
\begin{equation}
    \rho_{12}^+(\psi)= \frac{ (\ket{\psi}\bra{\psi} \otimes \mathbb{I})\rho_{12}^+ (\ket{\psi}\bra{\psi} \otimes \mathbb{I})} {\Tr [(\ket{\psi}\bra{\psi} \otimes \mathbb{I})\rho_{12}^+ (\ket{\psi}\bra{\psi} \otimes \mathbb{I}) ] },
\end{equation}
\begin{equation}
    \rho_{12}^+(\bar{ \psi} )= \frac{ (\ket{ \bar{ \psi} }\bra{ \bar{ \psi} } \otimes \mathbb{I})\rho_{12}^+ (\ket{ \bar{ \psi} }\bra{\bar{ \psi} } \otimes \mathbb{I})}{\Tr [(\ket{ \bar{ \psi}}\bra{\bar{ \psi}} \otimes \mathbb{I})\rho_{12}^+ (\ket{\bar{ \psi} }\bra{\bar{ \psi} } \otimes \mathbb{I}) ]}.
\end{equation}
 After Alice communicates her measurement outcome, the conditional state of the second qubit is given by
\begin{align}
  \rho_{2}^+(\psi) &= \rho_{2}^+(\bar{ \psi} ) \notag\\&= \frac{1}{2+k} \splitfrac{ \{ \mathbb{I}+ k\{ \abs{\beta}^2 \ket{0}\bra{0} - \alpha\beta^* \ket{0}\bra{1}}{ -\alpha^* \beta \ket{1} \bra{0} + \abs{\alpha}^2 \ket{1}\bra{1} \} }
\label{eqn A1}
\end{align}
We calculate the coherence of the above state in accordance with the $l_1$ norm definition of coherence, $C(\rho_{2}^+(\psi))= \frac{\abs{k} 2 \abs{\alpha} \abs{\beta}}{\abs{2+k}}$. From this equation, we infer that coherence of the state given in (\ref{eqn A1}) is one-fifth of $C(\psi)$ for $p=0.5$. Quite counter-intuitively, we find that complete decoherence of the state did not take place as in the case of completely depolarizing channel. Interestingly, an advantage can be gained with respect to coherence as a result of superposition of a noisy channel with another copy of itself. Even though there is no entanglement shared between Alice and Bob, there is nonzero quantum discord present across $\rho_{12}^+$. The non-zero discord helps to create non-zero quantum coherence with the help of indefinite causal order.

\begin{figure}[h]
\centering
\includegraphics[width=0.995\linewidth]{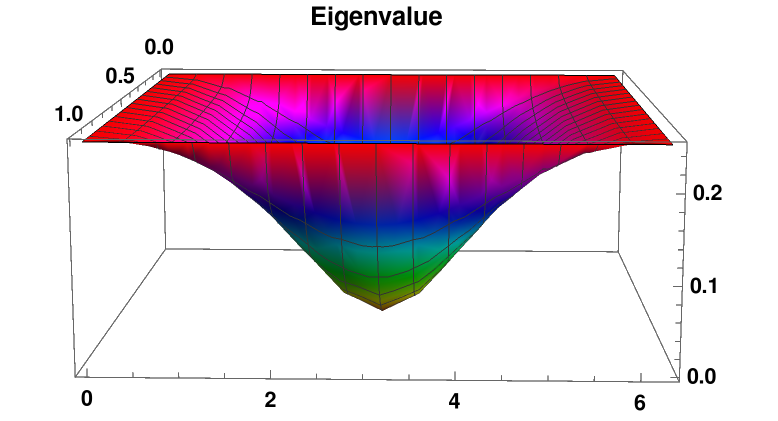}
  \caption{Plot of the second eigenvalue of the partially transposed density matrix of $\rho_{12}^+$ as in Eq.(15).}
  \label{fig::2eig}
\end{figure}

\subsection{One unitary operator and a completely depolarizing channel} 

\begin{figure}[h]
\centering
\includegraphics[width=0.995\linewidth]{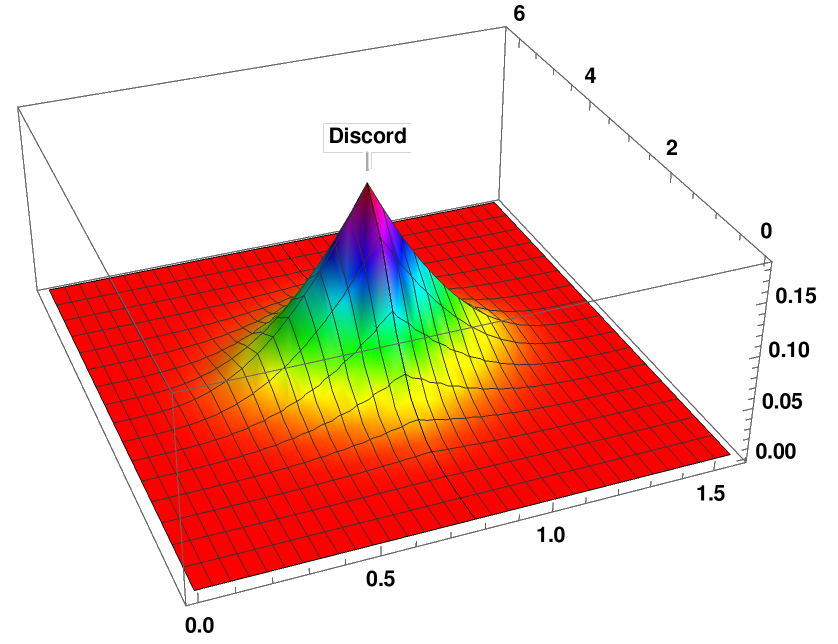}
  \caption{Quantum discord of the state $\rho_{12}^+$ that survives under superposition of noisy channels.}
   \label{fig::discord}
\end{figure}

In this section, we consider a noisy scenario where we have the superposition of two channels acting on the half of the entangled pair. Let this act on Bob's side. The first channel involved in the indefinite causal order is completely depolarizing and the second one is a unitary channel $U$. The shared state after the action of such noise is given by
\begin{equation}
\begin{split}
    \rho^{out}_{12C}= \frac{\mathbb{I}}{2} \otimes \{ \frac{\mathbb{I}}{2} \otimes p \ket{0}\bra{0} + U^\dagger \frac{\cos{\frac{\gamma}{2}}}{2} \otimes k \ket{0}\bra{1}  \\ + U \frac{\cos{\frac{\gamma}{2}}}{2} \otimes k \ket{1}\bra{0}+ \frac{\mathbb{I}}{2} \otimes (1-p) \ket{1}\bra{1} \} \\ -  \frac{k}{4} \{ \ket{0} \bra{1} \otimes (-\sin{\frac{\gamma}{2}} e^{-i\delta} U^\dagger) \otimes \ket{0}\bra{1} \\ + \ket{0} \bra{1} \otimes ( \sin{\frac{\gamma}{2}} e^{-i\delta} U) \otimes \ket{1}\bra{0} \} \\ -  \frac{k}{4} \{ \ket{1} \bra{0} \otimes (\sin{\frac{\gamma}{2}} e^{i\delta} U^\dagger) \otimes \ket{0}\bra{1} \\ - \ket{1} \bra{0} \otimes ( \sin{\frac{\gamma}{2}} e^{i\delta} U) \otimes \ket{1}\bra{0} \}, 
\end{split}    
\end{equation}
where $U$ is of the form
\begin{equation}
U = \begin{bmatrix}\cos{\frac{\gamma}{2}} & -\sin{\frac{\gamma}{2}}e^{-i\delta} \\\sin{\frac{\gamma}{2}}e^{i\delta} & \cos{\frac{\gamma}{2}} \end{bmatrix}.
\end{equation}
After performing measurement on the control qubit in the basis $\{ \ket{+}, \ket{-} \}$, the conditional state (for the outcome $+$ ) shared between Alice and Bob is given by
\begin{equation}
    \begin{split}
        \rho_{12}^+= \frac{1}{2} \Big[ \frac{\mathbb{I}}{2} \otimes \frac{\mathbb{I}}{2} + \frac{k}{2} \{\frac{\mathbb{I}}{2} \otimes{\cos{\frac{\gamma}{2}}} U^\dagger 
        + \frac{\mathbb{I}}{2} \otimes {\cos{\frac{\gamma}{2}}} U \\ +\frac{1}{4} \ket{0} \bra{1} \otimes \sin{\frac{\gamma}{2}} e^{-i\delta} U^\dagger - \frac{1}{4} \ket{0} \bra{1} \otimes \sin{\frac{\gamma}{2}} e^{-i\delta} U  \\- \frac{1}{4} \ket{1} \bra{0} \otimes \sin{\frac{\gamma}{2}} e^{i\delta} U^\dagger  + \frac{1}{4} \ket{1} \bra{0} \otimes \sin{\frac{\gamma}{2}} e^{i\delta} U\}\Big].
    \end{split}
\end{equation}
By the PPT criterion of entanglement, we infer that the state shared between Alice and Bob is also not entangled. The eigenvalues of the partially transposed state $\rho_{12}^+$ are given by two copies of each of 
$\frac{1+2k}{4(1+k(1+\cos{\gamma}))},
\frac{1+2k\cos{\gamma}}{4(1+k(1+\cos{\gamma}))}\nonumber
$.
The first one is manifestly positive and the second one also is positive as evident from the Fig.(\ref{fig::2eig}). Therefore, by the PPT criteria the state is not entangled. Next, we check whether the state has some nonzero amount of quantum discord and find that indeed it has some amount of quantum discord. 
For the range $0\leq\delta\leq\frac{\pi}{4}$, we have the following expression for quantum discord
\begin{widetext}
{\small{\begin{align}
   D(\rho_{12}^+)=\frac{1}{4(1+(k+1)\cos{\gamma})}\Big [ (1+2k)\log_2(\frac{(1+2k)}{(1+(k+1)\cos{\gamma})})
  +  (1+2k\cos{\gamma})\log_2(\frac{(1+2k\cos{\gamma})}{(1+(k+1)\cos{\gamma})}) \nonumber\\
   - 2(1+k(1+\cos{\gamma}-2\cos^2{\delta}\sin^2{\frac{\gamma}{2}}))\log_2(1-\frac{2k\cos^2{\delta}\sin^2{\frac{\gamma}{2}}}{(1+(k+1)\cos{\gamma})}) 
    - 2(1+k(1+\cos{\gamma}+2\cos^2{\delta}\sin^2{\frac{\gamma}{2}}))\log_2(1+\frac{2k\cos^2{\delta}\sin^2{\frac{\gamma}{2}}}{(1+(k+1)\cos{\gamma})}) \\ \nonumber
    +  2(1+k(1+\cos{\gamma}+2\cos{2\delta}\sin^2{\frac{\gamma}{2}}))\log_2(1+\frac{2k\cos{2\delta}\sin^2{\frac{\gamma}{2}}}{(1+(k+1)\cos{\gamma})})
      + 2(1+k(1+\cos{\gamma}-2\cos{2\delta}\sin^2{\frac{\gamma}{2}}))\log_2(1-\frac{2k\cos{2\delta}\sin^2{\frac{\gamma}{2}}}{(1+(k+1)\cos{\gamma})}) \nonumber
    \Big].
\end{align}}}

\end{widetext}
For the range $\frac{\pi}{4}\leq\delta\leq\frac{\pi}{2}$, we have the following expression for quantum discord
\begin{widetext}
{\small{\begin{align}\nonumber
    D(\rho_{12}^+)=\frac{1}{4(1+(k+1)\cos{\gamma})}\Big [ (1+2k)\log_2(\frac{(1+2k)}{(1+(k+1)\cos{\gamma})}) 
    + (1+2k\cos{\gamma})\log_2(\frac{(1+2k\cos{\gamma})}{(1+(k+1)\cos{\gamma})}) \nonumber\\
   -  2(1+k(1+\cos{\gamma}-2\sin^2{\delta}\sin^2{\frac{\gamma}{2}}))\log_2(1-\frac{2k\sin^2{\delta}\sin^2{\frac{\gamma}{2}}}{(1+(k+1)\cos{\gamma})})
    - 2(1+k(1+\cos{\gamma}+2\sin^2{\delta}\sin^2{\frac{\gamma}{2}}))\log_2(1+\frac{2k\sin^2{\delta}\sin^2{\frac{\gamma}{2}}}{(1+(k+1)\cos{\gamma})})\\ \nonumber
     + 2(1+k(1+\cos{\gamma}+2\cos{2\delta}\sin^2{\frac{\gamma}{2}}))\log_2(1+\frac{2k\cos{2\delta}\sin^2{\frac{\gamma}{2}}}{(1+(k+1)\cos{\gamma})})
      + 2(1+k(1+\cos{\gamma}-2\cos{2\delta}\sin^2{\frac{\gamma}{2}}))\log_2(1-\frac{2k\cos{2\delta}\sin^2{\frac{\gamma}{2}}}{(1+(k+1)\cos{\gamma})}) \nonumber
    \Big].
\end{align}}}
\end{widetext}
For a chosen value $p=0.25$, the plot of the quantum discord is given in Fig (\ref{fig::discord}).
Now, the question that we raise here is that whether it is possible to create perfect quantum coherence with this shared state in hand. To check this, we follow the following steps. 
Initially, we allow Alice to perform measurement in basis $\{\ket{\psi}, \ket{ \bar{\psi} } \}$. After sending the result of measurement outcome and tracing out the first party, 
we look at the state at Bob's side. The conditional state at the remote location is given by
\begin{align}
  \rho_{2}^+(\psi) &= \rho_{2}^+(\bar{\psi}) \notag\\ &= \frac{\splitfrac{\mathbb{I} +k\cos{\frac{\gamma}{2}}(U^\dagger+U )
 +k}{\sin{\frac{\gamma}{2}}(U^\dagger-U )(e^{-i\delta}\alpha^*\beta-e^{i\delta}\alpha\beta^*)}}{2+4k\cos^2{\frac{\gamma}{2}}}.
\label{eqn: A} 
\end{align}
Using the $l_1$ norm definition of coherence, we find that the coherence of the above shared state is given by 
\begin{equation}
    C(\rho_{2}^+(\psi))= \frac{2k\abs{\sin{\frac{\gamma}{2}}}^2\abs{(e^{-i\delta}\alpha^*\beta-e^{i\delta}\alpha\beta^*)}}{\abs{1+2k\cos^2{\frac{\gamma}{2}}}}.
\end{equation}
Now, one may ask if there are conditions on the parameters $p$, $\gamma$ and $\delta$ such that the coherence of the state created at Bob's lab is exactly $C(\psi) = 2\abs{\alpha}\abs{\beta}$. The conditions obtained with $p$ being equal to $0.5$ are given by
  $ \gamma = (2n+1)\frac{\pi}{2}$ and 
    $\delta = \frac{\pi}{2}-\phi$.
If Alice and Bob share maximally entangled states and this type of noise acts on half of the EPR pair, then exactly $C(\psi)$ amount of coherence can be created for the above stated conditions for $p$, $\gamma$ and $\delta$. Hence, remote creation of coherence is taking place in this scenario and indefinite causal order is undoubtedly playing a major role in this creation. Notice that without the use of indefinite causal order, if two channels act one after the other, then it is not possible to create a non-zero coherence at a remote location.

\subsection{Two partially depolarizing channels}
In this section we show that the application of indefinite causal order of two partially depolarizing channel also creates more coherence than that of series application of two partially depolarizing channels. The action of partially depolarizing channel
$\rho\rightarrow \sum_i K_i\rho K_i^\dagger$
where the Kraus operators are defined as follows
$K_0=\sqrt{1-\frac{3q}{4}}\mathbb{I}, K_1=\frac{\sqrt{q}}{2}\sigma_x,K_2=\frac{\sqrt{q}}{2}\sigma_y,K_3=\frac{\sqrt{q}}{2}\sigma_z.$
Therefore, we have the following output state under the superposition of partially depolarizing channels 
$  \rho^{out}_{12c}=\sum_{ij}(\mathbb{I}\otimes W_{ij})(|\Psi^-\rangle\langle\Psi^-|\otimes \rho_c)(\mathbb{I}\otimes W_{ij}^{\dagger})$,
where $W_{ij}$ will act on half of the EPR pair and the control qubit. In the above equation we have
$\rho_c=|\Psi_c\rangle\langle\Psi_c|$, where $|\Psi_c\rangle=\sqrt{p}|0\rangle+\sqrt{1-p}|1\rangle$
and $W_{ij}=K_iK_j\otimes |0\rangle\langle 0|+ K_jK_i\otimes|1\rangle\langle 1| $.
 
Now, if measurement if performed on the control qubit in the $\{ \ket{+}, \ket{-}\}$ basis, then the conditional state shared between Alice and Bob (for the measurement outcome, for e.g., $\ket{+}$ ) is given by $\rho_{12}^{\pm}=\langle\pm|\rho^{out}_{12c}|\pm\rangle$, where $|\pm\rangle=\frac{1}{\sqrt{2}}(|0\rangle\pm|1\rangle)$. Next, Alice performs measurement in $\{\ket{\psi}, \ket{ \bar{\psi} } \}$ basis. Depending on the measurement outcome, we look at the state of Bob's qubit.
By tracing out the first qubit, the state of Bob's qubit is given by 
\begin{widetext}
\begin{equation}
  \rho_{2}^+(\psi) = \begin{bmatrix}\frac{-4\beta^2-4q(\alpha^2-\beta^2)+2q^2(\alpha^2-\beta^2)+(-4\beta^2-4q(\alpha^2-\beta^2)+2q^2(4\alpha^2-\beta^2))k}{(-4+2(-4+3q^2)k)} & \frac{e^{-i\phi}\alpha\beta(2(1-q)^2+(4-8q+5q^2)\sin{\theta}\cos{\theta})}{(-2-(4-3q^2)k)} 
    \\\frac{e^{i\phi}\alpha\beta(2(1-q)^2+(4-8q+5q^2)k)}{(-2-(4-3q^2)k)}  & \frac{-2(2-2q+q^2)\alpha^2-2(2-q)q\beta^2-2((2-q)^2\alpha^2-4(1-q)q\beta^2)k}{(-4-2(4-3q^2)k)}
      \end{bmatrix}  
\end{equation}
\end{widetext}
The coherence of the above state is given as the following
 \begin{equation}
  C(\rho_2^+) = |\frac{2\alpha\beta(2(1-q)^2+(4-8q+5q^2)k)}{(-2-(4-3q^2)k)}|
 \end{equation}

 If there is no causal order and we apply the partial depolarising channel on one-half of the entangled pair, then after following the remote creation of coherence protocol, we find that the coherence at Bob's location will be $2|\alpha\beta(1-q)^2|$
Since the coherence of the initial state is $C(\psi) = 2|\alpha\beta|$, the fraction of coherence compared to the coherence of the initial state, in the absence of indefinite causal order is given by $ (1-q)^2$. 
Similarly, the fraction of coherence compared to the coherence of the initial state, created by the use of indefinite causal order is 
given by $ R_c(\rho_2^+)=\frac{C(\rho_2^+)}{2|\alpha\beta|}$. Therefore, we have the following expressions for the fraction of coherence in the 
presence and absence of indefinite causal order as given by
  \begin{align*}
      R_c(\rho_2^+) = &|\frac{(2(1-q)^2+(4-8q+5q^2)k)}{(-2-(4-3q^2)k)}| ~~\mathrm{and}
      ~~|(1-q)^2|
  \end{align*}
    \begin{figure}[h]
\centering
\includegraphics[width=1.2\linewidth]{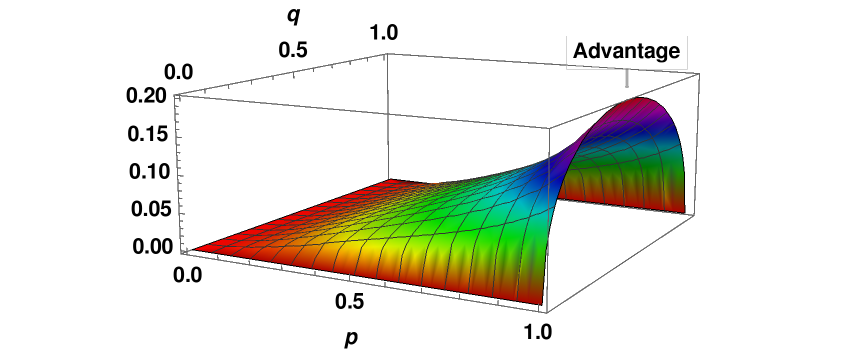}
  \caption{Advantage of remote creation of quantum coherence via indefinite causal order, conditioned on $|+\rangle$ state.}
  \label{fig::adv}
  \end{figure}
 We plot the difference $R_c(\rho_2^+)-|(1-q)^2|$ in Fig (\ref{fig::adv}) to see if we gain an advantage by using indefinite causal order. We see from Fig (\ref{fig::adv}) that in most cases we have a positive value showing that indeed there is an advantage of remote creation of quantum coherence by using indefinite causal order. 

\section{Conclusions and Future Directions}
To summarise, we described a method to create quantum coherence at a remote location using shared entanglement, local operation and classical communication. 
If Alice and Bob share a maximally entangled state, then Alice can create quantum coherence of any arbitrary qubit state at Bob's location perfectly by communicating one classical bit.
This is in sharp contrast to remote state preparation, where Alice can prepare a qubit chosen from special ensembles, i.e., a qubit chosen from equatorial or polar great circles. If the shared state undergoes a noisy channel, then typically, it losses quantum entanglement and other quantum correlations. However, if two noisy channels are in the superposition of indefinite causal order, then, even though the entanglement is lost, we find that the presence of tiny amount of quantum discord can help to create a non-zero quantum coherence. 
This shows that the creation of quantum coherence at a remote location can be enhanced by the usage of indefinite causal order. Specifically,
we have shown that, while a completely decohering channel does not preserve any form of coherence, nevertheless, the use of indefinite causal order between two such decohering channels is able to create quantum coherence at a remote location via the use of maximally entangled state. 
We also explored the creation of quantum coherence for the superposition of one completely decohering channel and one unitary operator.
Interestingly, in the case of superposition of completely randomising channel and unitary operator, it is possible to create complete coherence at a remote location for some choice of the unitary operators.
Furthermore, we have shown an advantage of creating coherence via indefinite causal order of two partially polarizing channel vs without the use of indefinite causal order. These results may have important application in the creation of coherence in noisy scenarios.
The mechanism investigated here may provide a better understanding on how to maintain coherence under extreme noisy channels. In future, we will explore the possibility of preserving quantum coherence in the hot environments with the help of indefinite causal order. 

\vskip 1cm
\section*{Data Availability Statement }The datasets generated during and/or analysed during the current study are available from the corresponding author on reasonable request.
\vskip 1cm
\section*{Acknowledgements} J. Kaur acknowledges financial support during her visit to HRI, Allahabad from Dec 2019 to March 2020. 
S. Bagchi acknowledges funding by PBC Fellowship at Tel Aviv university. A. K. Pati acknowledges funding from the J C Bose Research Grant from
the Department of Science and Technology (DST), India.

\end{document}